\documentclass[aps, reprint, prl, twocolumn, superscriptaddress,amsmath,amssymb,noeprint]{revtex4-2}
\usepackage{times}
\usepackage{amsmath}
\usepackage{amssymb}
\usepackage{xfrac}
\usepackage{dsfont}
\usepackage{graphicx}
\usepackage{float}
\usepackage{braket}
\usepackage{bbold}
\usepackage[normalem]{ulem}
\usepackage{bm}
\usepackage{bbm}
\usepackage{color}
\usepackage{xcolor}

\usepackage{pdfpages}
\usepackage{pgffor}

\makeatletter
\AtBeginDocument{\let\LS@rot\@undefined}
\makeatother

\usepackage[colorlinks=true, urlcolor=blue, linkcolor=blue, citecolor=blue]{hyperref}
\usepackage{siunitx}

\newcommand{\berkeleyphy}{Department of Physics, University of California, Berkeley, California 94720}
\newcommand{\CIQC}{Challenge Institute for Quantum Computation, University of California, Berkeley, California 94720}
\newcommand{\LBL}{Materials Sciences Division, Lawrence Berkeley National Laboratory, Berkeley, California 94720}
\newcommand{\columbia}{Department of Physics, Columbia University, New York, NY 10027}

\begin{document}

\title{Super-radiant and Sub-radiant Cavity Scattering by Atom Arrays}

\author{Zhenjie Yan}
\affiliation{\berkeleyphy}
\affiliation{\CIQC}

\author{Jacquelyn Ho}
\affiliation{\berkeleyphy}
\affiliation{\CIQC}

\author{Yue-Hui Lu}
\affiliation{\berkeleyphy}
\affiliation{\CIQC}

\author{Stuart J. Masson}
\affiliation{\columbia}

\author{Ana Asenjo-Garcia}
\affiliation{\columbia}

\author{Dan M. Stamper-Kurn}
\email[]{dmsk@berkeley.edu}
\affiliation{\berkeleyphy}
\affiliation{\CIQC}
\affiliation{\LBL}
\begin{abstract}

We realize collective enhancement and suppression of light scattered by an array of tweezer-trapped $^{87}$Rb atoms positioned within a strongly coupled Fabry-P\'{e}rot optical cavity.  We illuminate the array with light directed transverse to the cavity axis, in the low saturation regime, and detect photons scattered into the cavity.  
For an array with integer-optical-wavelength spacing each atom scatters light into the cavity with nearly identical scattering amplitude, leading to an observed $N^2$ scaling of cavity photon number as the atom number increases stepwise from $N=1$ to $N=8$.
By contrast, for an array with half-integer-wavelength spacing, destructive interference of scattering amplitudes yields a non-monotonic, sub-radiant cavity intensity versus $N$.  By analyzing the polarization of light emitted from the cavity, we find that Rayleigh scattering can be collectively enhanced or suppressed with respect to Raman scattering.  We observe also that atom-induced shifts and broadenings of the cavity resonance are precisely tuned by varying the atom number and positions.   Altogether, tweezer arrays provide exquisite control of atomic cavity QED spanning from the single- to the many-body regime.

\end{abstract}

\maketitle

As highlighted by Dicke's seminal work on super- and sub-radiance \cite{Dicke1954}, the interaction of multiple emitters with a quantum mode of light differs from that of the emitters individually.
Collective ``super-radiant'' (or ``bright'') states, resulting from constructive interference, give rise to an enhanced emission rate per excitation, which grows with the number of emitters.
Conversely, ``sub-radiant'' (or ``dark'') states arise from destructive interference, leading to a suppression or complete cancellation of photon emission.

In the case of extended samples, with emitters distributed over distances longer than the emitted optical wavelength, the observation and control of super- and sub-radiance depends critically on the exact spatial distribution of the emitters.  For example, the precise structure of mesoscopic samples \cite{Clemens2003,Masson2020,Scully2006} or periodic emitter arrays \cite{Masson2022,Tamura2020,Rui2020} controls whether their collective emission is enhanced or suppressed, or directed into single or multiple optical modes.

In cavity quantum electrodyamics (QED), in which each of multiple emitters couples strongly to a single-mode cavity field, the properties of (and the access to) the bright and dark manifolds depend strongly on the spatial positions of emitters within the cavity.
Already for single emitters, controlling the position of atoms in a cavity advanced the field of cavity QED ~\cite{Guthohrlein2001,Nussmann2005,Reiserer2013a,Thompson2013a}. Basic effects of few-body cavity QED have been illustrated by the controlled placement of two atoms~\cite{Reimann2015,Neuzner2016} or ions~\cite{Casabone2015,Begley2016} within resonant cavities. 
Experiments on superconducting quantum circuits have extended studies of collective emission into microwave cavities and waveguides to as many as 10 qubits~\cite{Majer2007,VanLoo2013,Mirhosseini2019,Kannan2020,Wang2020a}.

\begin{figure}
\centering
\includegraphics[width=3.375in]{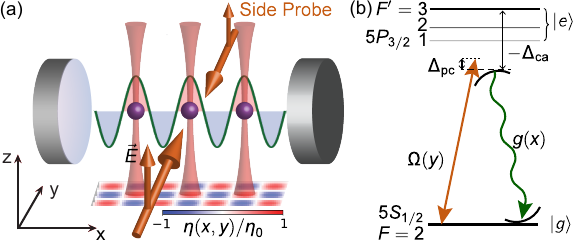}
\caption{Schematic of the experimental setup. (a) Atoms are driven by a pair of counter propagating side probe beams. 
The two-photon scattering amplitude $\eta(x,y)$ from the probe beam into the cavity, by an atom located at $(x, y)$, exhibits a two-dimensional checkerboard pattern.
(b) The probe light (orange) and the cavity photons (green) couple the $F=2$ hyperfine states of the $5S_{1/2}$ atomic ground state with the $5P_{3/2}$ excited states.
Three hyperfine manifolds $F^\prime=3,2,1$ of the excited states contribute to the probe-cavity scattering.
}
\label{fig:M1}
\end{figure}

In this work, we employ deterministically loaded atom tweezer arrays~\cite{Endres2016,Barredo2016}, a powerful new platform for quantum simulation~\cite{Browaeys2020}, metrology~\cite{Madjarov2019,Young2020}, and information processing~\cite{Kaufman2021}, to advance atomic cavity QED from the few- to the many-body regime while preserving precise control over the cavity interaction of each individual atom.
With such a tweezer array, we place a fixed number of $^{87}$Rb atoms at fully controlled positions along the axis of a strongly coupled Fabry-P\'{e}rot optical resonator~\cite{Liu2022,comment_zhang} [Fig.~\ref{fig:M1}(a)].
This method improves upon the incomplete control of atom number, position, and motion in previous approaches used in trapped-atom and -ion cavity QED experiments~\cite{Reiserer2015,Mivehvar2021}. 
By driving this emitter array with light propagating transverse to the cavity axis while monitoring the cavity emission, we prepare both super- and sub-radiant (single-excitation) states. We characterize these states via the number of photons present in the cavity, which scales as $N^2$ for super-radiant states and sublinearly for sub-radiant ones.

\begin{figure*}
\centering
\includegraphics[width=\textwidth]{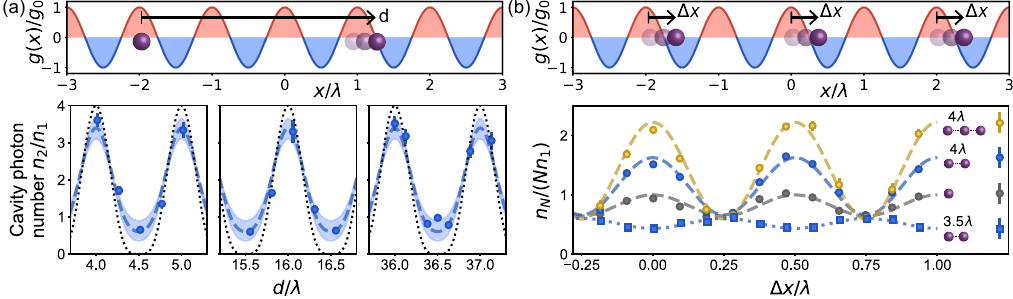}
\caption{
(a) Cavity scattering by a two-atom array. One atom is placed at a cavity antinode, and scatters with steady-state cavity photon number $n_1$ on its own. Adding a second atom, at a distance $d$ from the first, changes the cavity photon number to $n_2$. 
Expected ratios $n_2/n_1$ for perfectly localized atoms (dotted black line) and atoms with $\sigma=100(14)$ nm position fluctuations (dashed blue line, uncertainty in $\sigma$ indicated by shaded area).
(b) Normalized cavity photon number $n_N/(N n_1)$ generated by $N$-atom arrays with fixed spacing, displaced by $\Delta x$ from a cavity antinode. 
The array has $N=1$ (gray), $2$ (blue), $3$ (yellow) atoms, and a spacing of $4\lambda$ (circles) or $3.5\lambda$ (squares).
The dashed and dotted lines are cosinusoidal fits with a period of $\lambda/2$. 
All data are taken with $\Delta_\mathrm{pc} = 0$.
}
\label{fig:M2}
\end{figure*}

We employ an in-vacuum near-concentric Fabry-P\'{e}rot optical cavity with a TEM$_{00}$ mode whose frequency ($\omega_\mathrm{c}$)  lies near the transition frequency ($\omega_\mathrm{a}$) of the $^{87}$Rb $F=2 \rightarrow F^\prime=3$ $D_2$ transition (wavelength $\lambda = \SI{780}{\nano\meter}$) ~\cite{Deist2022a,Deist2022b}.
Near the cavity center, the coupling amplitude of a single $^{87}$Rb atom to this cavity mode varies as $g(x) = g_0 \cos kx$; here, $g_0 = 2 \pi \times 3.1$ MHz  (on the cycling transition) and $k = 2 \pi/\lambda$.  Given the atomic and cavity resonance half-linewidths of $\gamma = 2 \pi \times 3.0$ MHz and $\kappa = 2 \pi \times 0.53$ MHz, respectively, the cavity achieves the strong coupling condition with single-atom cooperativity $C = g_0^2/(2 \kappa \gamma) = 3.0$~\cite{Kimble1998}.

A one-dimensional array of optical tweezers is formed by laser beams with a wavelength of \SI{808}{\nano \meter} sent transversely to the cavity through a high numerical-aperture imaging system.
Pre-cooled and optically trapped $^{87}$Rb atoms are loaded into as many as 16 tweezers, detected through fluorescence imaging, and then sorted into regularly spaced arrays of $N=1$ to $8$ atoms~\cite{suppmat}\nocite{Madjarov2021a}. The total length of these arrays is much smaller than the $\sim \SI{1}{\milli\meter}$ cavity Rayleigh range. The array is aligned to place atoms at the radial center of the cavity.  Piezo-controlled mirrors and an acousto-optical deflector are used to position the array with nanometer-scale precision along the cavity axis.

Our experiments focus on the response of this atom array when it is driven optically.  
For this, we illuminate the array with a transverse standing-wave of monochromatic probe light, see Fig.~\ref{fig:M1}(a).
The probe is linearly polarized along the $z$-axis (chosen as our quantization axis), perpendicular to the cavity, and drives each atom with a spatially dependent Rabi frequency $\Omega(y)=\Omega_0 \cos{k y}$.
The standing-wave configuration balances photon recoil momentum, allows for positional calibration along the $y$-axis, and reduces incoherence in photon emission caused by atomic thermal motion~\cite{suppmat}.
We employ a weak probe amplitude to ensure that the probability of having more than one atom excited simultaneously is negligible.
All tweezer traps are centered at $y=0$. 
The probe frequency $\omega_\mathrm{p}$ operates at a small detuning  $\Delta_\mathrm{pc}=\omega_\mathrm{p}-\omega_\mathrm{c}$ from the cavity resonance [Fig.~\ref{fig:M1}(b)].  Prior to probe illumination, the tweezer-trapped atoms are polarization-gradient cooled and prepared in the $F=2$ ground hyperfine manifold, without control of their Zeeman state.  Probe light scattered by the array into the cavity, and thence through one mirror of our one-sided cavity, is detected by a single-photon counting module.

To demonstrate the strong sensitivity of collective cavity scattering on the exact positioning of the scatterers, we first examine an array of just two atoms.  As illustrated in Fig.\ \ref{fig:M2}(a), we first position one atom at an antinode of the cavity field.  With the cavity at a large detuning $\Delta_\mathrm{ca} = \omega_\mathrm{c} - \omega_\mathrm{a} = - 2 \pi \times 507$ MHz --- a specific ``magic'' value chosen to suppress fluctuations from internal-state dynamics, as described below --- we denote the steady-state cavity photon number generated by this single atom as $n_1$.  We then add a second atom at a variable distance $d$ from the first, and record the cavity photon number produced by the atom pair as $n_2$.  At integer-wavelength separation ($d = m \lambda$ with $m$ an integer), we observe super-radiant light scattering, where the total cavity emission rate is greater than that of two individual atoms ($n_2 > 2 n_1$).  At half-integer-wavelength separation ($d = (m+1/2) \lambda$), we observe sub-radiant light scattering ($n_2 < 2 n_1$).

To account for the observed behavior, let us consider that each atom $i$, positioned at location $(x_i, y_i)$, serves as a source for light in the cavity, with elastic scattering amplitude $\eta(x_i, y_i) = g(x_i) \Omega(y_i)/2 \Delta_\mathrm{ca} \equiv \eta_0 \cos kx_i \cos ky_i$, which we obtain by treating the atom as a two-level emitter and adiabatically eliminating its excited state.  The scattering amplitudes from all atoms add coherently.  The steady-state cavity photon number scattered by $N$ atoms is then $n_N = |\bar{a}|^2 $, where the expectation value of the cavity-field amplitude is given, following a semi-classical treatment~\cite{Domokos2002,Tanji-Suzuki2011,suppmat} in the dispersive coupling regime $|\Delta_\mathrm{ca}| \gg \{\Omega_0,g_0,\gamma\}$, as
\begin{align}
\bar{a}=\frac{\sum_{i}\eta(x_i,y_i)}{[\Delta_\mathrm{pc}-\sum_{i} g^2(x_i)/\Delta_\mathrm{ca} ] + i[\kappa + \sum_{i} \gamma g^2(x_i)/{\Delta_\mathrm{ca}^2}  ]}.
\label{eq:CavityField}
\end{align}
Here, we have accounted for the atom-induced dispersive shift $\sum_i g^2(x_i)/\Delta_\mathrm{ca}$ and absorptive broadening $\sum_i \gamma g^2(x_i)/\Delta_\mathrm{ca}^2$ of the cavity resonance.

At large $|\Delta_\mathrm{ca}|$, where we can neglect atom-induced modifications of the cavity resonance, and with $\Delta_\mathrm{pc}=0$, the cavity photon number varies simply as $n_N = |\sum_i \eta(x_i, y_i)|^2/\kappa^2$.  For two atoms, with the first situated exactly at the antinode and the second at an exact axial distance $d$, one then expects $n_2/n_1 = [1+\cos kd]^2$, with limiting values $n_2/n_1 = 4$ from constructive interference at integer-wavelength separation, and $n_2/n_1 = 0$ from destructive interference at half-integer-wavelength separation.  In our experiment, uncorrelated fluctuations in $\eta$, deriving from thermal position fluctuations of the two atoms within their tweezer traps, constrain these limiting values to $n_2/n_1 = 2 \left( 1 \pm D  \right)$, where the ratio $D = |\langle \eta\rangle|^2/ \langle |\eta|^2\rangle$ is the Debye-Waller factor and $\langle \rangle$ denotes an average over the position distribution of a single trapped atom.  The data in Fig.\ \ref{fig:M2}(a) are consistent with root mean square variations of $\sigma = 100(14)$ nm in both the $x$ and $y$ directions of motion, with  $\sigma$ determined independently by measuring light scattering from a single atom \cite{suppmat}.  In future work, $\sigma$ can be reduced further through bursts of dark-state cooling ~\cite{Kaufman2012,AngOngA2022,Brown2019}.  Strong contrast between constructive and destructive interference is retained even at large atomic separations -- as far as  $d\simeq \SI{30}{\micro\meter}$.

The photon scattering rate also depends on the positions of the atoms with respect to the standing-wave cavity mode. 
We illustrate this dependence by measuring light scattering from atom arrays with fixed integer- or half-integer-wavelength spacing while translating these arrays altogether by $\Delta x$ from a cavity antinode [Fig.\ \ref{fig:M2}(b)].  Again, we observe either super- or sub-radiance when the array is aligned onto cavity antinodes.  In contrast, when either type of array is aligned to cavity nodes, we observe scattering that scales linearly with $N$.  Here, position fluctuations cause the scattering amplitude from each atom to vary between positive and negative values with equal probability, producing a cavity field with finite variance but zero average amplitude.

\begin{figure}
\centering
\includegraphics[width=3.375in]{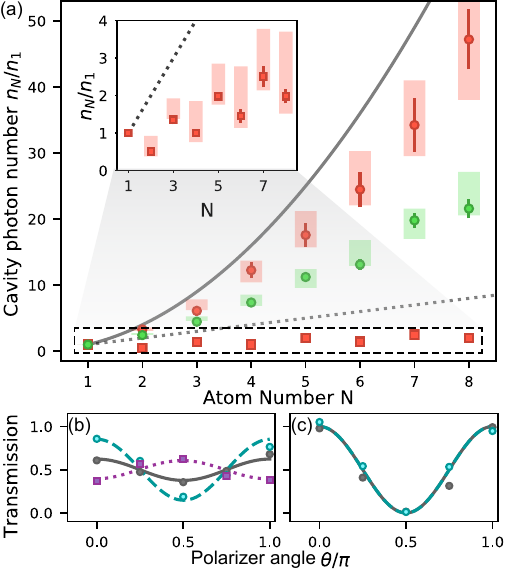}
\caption{
(a) Normalized cavity photon number ($n_N/n_1$) for scattering by $N$-atom arrays with either integer- (circles, $d=5.0 \lambda$) or half-integer- (squares, $d = 5.5 \lambda$) wavelength spacing, using cavity-atom detuning at either $\Delta_\mathrm{ca}=-2\pi \times 507$ MHz (red) or $-2 \pi \times 38$ MHz (green). 
Inset: zoomed-in data from half-integer-wavelength arrays.
The solid (dashed) black line corresponds to a quadratic $N^2$ (linear $N$) scaling.
The red shaded area is calculated from Eq.~\eqref{eq:N_atom_constructive} and \eqref{eq:N_atom_destructive}, while the green shaded area is the result of a Monte Carlo calculation assuming equal population among all $m_F$ states~\cite{suppmat}.
(b-c) Polarization analysis of cavity emission.  Relative transmission through a linear polarizer at an angle $\theta$ with respect to the $z$-axis at (b) $\Delta_\mathrm{ca}=-2\pi\times\SI{38}{\mega\hertz}$ and (c) $\Delta_\mathrm{ca}=-2 \pi \times \SI{507}{\mega\hertz}$.  Shown for scattering by a single atom (gray circles), or $N=8$ atoms at the constructive (cyan circles) or the destructive interference condition (purple squares). The corresponding curves show results of the Monte Carlo calculation.
}
\label{fig:M3}
\end{figure}

The super-radiant emission by atoms at the constructive interference condition is further enhanced by increasing the atom number [Fig.\ \ref{fig:M3}(a)].  At large $|\Delta_\mathrm{ca}|$ and on cavity resonance, the number of cavity photons emitted by an $N$-atom array is predicted to be
\begin{align}
n_N = \left[ N\left(\braket{|\eta|^2}-\lvert\braket{\eta}\rvert^2\right)+N^2 \lvert\braket{\eta}\rvert^2\right]/\kappa^2.
\label{eq:N_atom_constructive}
\end{align}
This expression consists of an incoherent part that scales linearly with $N$ and that vanishes if atoms are fully localized ($\sigma=0$), and a coherent part that scales quadratically as $N^2$.  Our measurements with $N$ ranging from 1 to 8 match well with this prediction, clearly exhibiting super-radiant scattering.

At the destructive interference condition, i.e.\ an atom array with half-integer-wavelength spacing, collective scattering is sub-radiant, falling below the linear scaling with $N$ expected for an incoherent sample.  Here, at similar probe and cavity settings, one expects the cavity photon number to vary as 
\begin{align}
n_N = \left[ N\left(\braket{|\eta^2|}-\lvert\braket{\eta}\rvert^2\right)+\frac{1-(-1)^N}{2} \lvert\braket{\eta}\rvert^2 \right]/\kappa^2.
\label{eq:N_atom_destructive}
\end{align}
While the incoherent scattering rate remains linear in $N$, the coherent photon scattering by pairs of atoms with opposite phase cancel out,  resulting in a total coherent contribution equal to that of either zero emitters (even $N$) or a single emitter (odd $N$).
The non-monotonic behavior observed in our experiment emerges as the coherent scattering rate exceeds the incoherent one: $\lvert\braket{\eta}\rvert^2>\braket{\lvert\eta^2\rvert}-\lvert\braket{\eta}\rvert^2$.

At smaller detuning $|\Delta_\mathrm{ca}|$, collective light scattering is affected by two additional effects.  First,  coherent scattering is degraded by polarization and intensity fluctuations arising from internal state dynamics of the multi-level $^{87}$Rb atom.  In our setup, incoherent Raman scattering causes each atom's magnetic quantum number $m_F$ to be distributed among all possible values.  For small $\Delta_\mathrm{ca} = - 2 \pi \times 38$ MHz, the amplitude $\eta$ for emitting $z$-polarized light, arising primarily from the near-resonant $F=2 \rightarrow F^\prime = 3$ transition, varies strongly with $m_F$.  This random variation further degrades the Debye-Waller factor, reducing the the collective enhancement of light scattering as shown in Fig.\ \ref{fig:M3}(a).

Furthermore, at this detuning, atoms can scatter the $z$-polarized probe light into cavity modes of two orthogonal polarizations. Scattering into the $z$-polarized cavity mode, i.e.\ Rayleigh scattering, does not change the spin state of the ground-state atoms.  The final spin state after Rayleigh scattering is independent of which atom scattered a photon.  Thus, the Rayleigh scattering rate is determined by the interference of the scattering amplitudes from all atoms, allowing for enhancement or suppression.  In contrast, scattering into the $y$-polarized cavity mode, i.e.\ Raman scattering, does change the spin state of the ground state. In our experiment, it is reasonable to assume the ground-state atoms are, at all times, in an incoherent mixture of spin states owing to imprecise initial state preparation and rapid state-collapse caused by incoherent light scattering. Given that the final spin states produced by each atom Raman scattering a photon are orthogonal, and absent coherence among initial spin states, the Raman scattering rates in our experiment add incoherently.
By operating in other regimes which allow for coherences among the ground states, collective Raman scattering can indeed be observed~\cite{Bohnet2012,Zhiqiang2017,Landini2018,Kroeze2018}.

In support of this description, we examine separately the cavity emission of $z$- and $y$-polarized light [Fig.~\ref{fig:M3}(b)].  At small detuning, and in comparison to the light polarization emitted by a single atom, the emission of $z$-polarized light (relative to $y$-polarized light) is enhanced by super-radiant scattering at integer-wavelength atomic spacing and suppressed by sub-radiant scattering at half-integer-wavelength atomic spacing.

For much of our work, we avoid these internal-state and polarization effects by operating at the aforementioned ``magic'' detuning of $\Delta_\mathrm{ca} = - 2\pi \times 507$ MHz~\cite{suppmat,paper_in_perp}.  Here, when accounting for transitions to all three accessible excited hyperfine states ($F^\prime = 1$, $2$, and $3$), the amplitude for scattering $z$-polarized light is nearly identical for all Zeeman states within the ground $F=2$ manifold.  Simultaneously, the rate for Raman scattering $y$-polarized light is nearly extinguished, as observed in the nearly pure $z$-polarization of light emitted from the cavity [Fig.\ \ref{fig:M3}(c)].
The identification of such a ``magic'' detuning for nearly all alkali species~\cite{suppmat,paper_in_perp}, which allows light scattering by an alkali atom to resemble closely that of just a two-level atom, should be beneficial to other quantum optics experiments using alkali gases.

\begin{figure}
\centering
\includegraphics[width=3.375in]{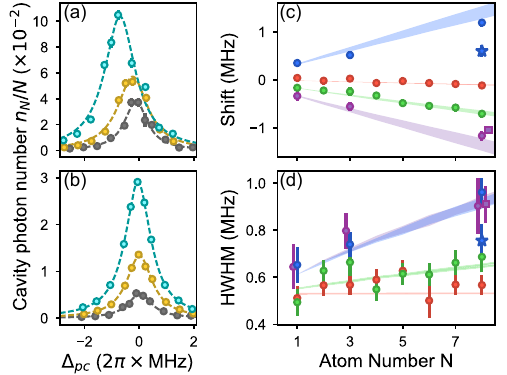}
\caption{
Cavity spectra measured with arrays of $N=1$ (black), $3$ (yellow), and $8$ (cyan) at (a) $\Delta_\mathrm{ca}=-2\pi\times\SI{38}{\mega\hertz}$ and (b) $-2\pi\times\SI{507}{\mega\hertz}$.
All atoms are aligned with cavity antinodes of the same phase. Lorentzian fits (dashed lines) are used to determine the peak positions and linewidths.
The cavity resonance (c) shifts and (d) half widths at half maximum (HWHM) as a function of atom number, measured with all atoms aligned with cavity antinodes of the same phase (circles), at detunings of $\Delta_\mathrm{ca}=-2\pi\times\SI{507}{\mega\hertz}$ (red), $-2\pi\times\SI{38}{\mega\hertz}$ (green), $-2\pi\times\SI{19}{\mega\hertz}$ (purple), and $2\pi\times\SI{19}{\mega\hertz}$ (blue).
The corresponding shaded areas are predictions from a Monte Carlo calculation~\cite{suppmat}.
Results are also shown for an array with half-integer-wavelength spacing at the cavity antinodes (squares), and an array with integer-wavelength spacing at the cavity nodes (stars).  
For clarity, some symbols in (c) and (d) are slightly offset in $N$.}
\label{fig:M4}
\end{figure}

Additionally, at small detuning, we observe significant modifications of the cavity resonance by the atomic array within. As shown in Fig.\ \ref{fig:M4}(a), we record $n_N$ for various arrays as a function of the detuning $\Delta_\mathrm{pc}$ of the probe from the empty-cavity resonance.  From the observed emission lineshapes, we extract the atom-induced cavity resonance shift and also the total cavity linewidth.  The measured spectral shifts and widths show a near-linear dependence on the atom number $N$ [Fig.~\ref{fig:M4}(c-d)], and agree well with a theoretical calculation accounting for both atomic position and $m_F$ fluctuations ~\cite{suppmat}.    
Owing to the $g^2(x_i)$ dependence of the cavity dispersion and absorption terms, the atom-induced cavity modification is the same for atoms at both integer- and half-integer-wavelength spacing.  
Notably, these modifications are reduced for atom arrays centered on the cavity field nodes, and such spatial dependence is the foundation of optomechanical coupling between the atomic motion and the cavity field~\cite{Baumann2010,Kohler2018}.
By comparison, the atom-induced cavity modifications are all negligible for large $|\Delta_\mathrm{ca}|$.

While the current work explores only the bottom of the Dicke ladder, where only one atom is excited, our platform enables study of many-excitation super- or sub-radiance~\cite{Dicke1954,Gross1982}, up to full or zero state inversion respectively, with unprecedented control.
The interplay between atomic quantum states, collective atom-cavity interaction, and optical emission properties can be fully explored~\cite{Hotter2023}.
Our setup would provide insight into key features of many-body decay in the presence of multiple decay channels (due to the multimode nature of the cavity), such as the scaling with atom number of the peak intensity and the time of peak emission, as well as large shot-to-shot fluctuations~\cite{Clemens2004,PineiroOrioli2022,Cardenas-Lopez2022} in the light polarization due to self-reinforcing feedback.

Our realization of a well-controlled many-body cavity QED system, compatible with single-atom control, offers a wide range of potential quantum applications.
When single-photon cavity emission is subradiantly suppressed, multi-photon emission becomes the dominant process, leading to a non-linear photon source~\cite{Habibian2011,Neuzner2016}.
Encoding quantum information in a symmetrically excited W state or related states~\cite{Cabello2002} can superradiantly speed up information exchange between matter and itinerant photons. This accelerates quantum communications and reduces the infidelity accumulated over the interaction time~\cite{Casabone2015,Zhong2017}.
The possibility of switching between super- and sub-radiant states, through either phase rotation on individual atoms or moving the tweezers, allows on-demand multi-photon storage and emission~\cite{Holzinger2022}. 

Finally, coherent long-range photon-mediated momentum~\cite{Mivehvar2021,Ritsch2013,Luo2023} or spin~\cite{Hosten2016b,Welte2018,Periwal2021a,Greve2022,Li2023a} exchanges between individual atoms can be facilitated through subsequent absorption of the cavity photons by the atoms and can be used for quantum simulation on the Dicke model~\cite{Kirton2019d}, entanglement generation~\cite{Li2022b}, or realization of quantum solvers~\cite{Torggler2019}.
Notably, atom arrays that combine long-range photon-mediated interactions with short-range Rydberg interactions could benefit from the advantages of both types of interactions with distinct ranges, connectivity, and time scales~\cite{Ramette2022,Huie2021a,Ocola2022}.

\begin{acknowledgments}
\textbf{Acknowledgment-} We acknowledge support from the AFOSR (Grant No.\ FA9550-1910328 and Young Investigator Prize Grant No.\ 21RT0751), from ARO through the MURI program (Grant No.\ W911NF-20-1-0136), from DARPA (Grant No.\ W911NF2010090), from the NSF (QLCI program through grant number OMA-2016245, and CAREER Award No.\ 2047380), and from the David and Lucile Packard Foundation.
J.H. acknowledges support from the National Defense Science and Engineering Graduate (NDSEG) fellowship.
We thank Erhan Saglamyurek for helpful discussions.
We also thank Jacopo De Santis, Florian Zacherl, and Nathan Song for their assistance in the lab.
\end{acknowledgments}
\bibliography{main.bib}

\begin{thebibliography}{69}
\bibitem [{pap()}]{paper_in_perp}%
  \BibitemOpen
  \href@noop {} {}\bibinfo {note} {The authors are currently preparing a
  separate manuscript on the general existence of such magic detuning in alkali
  atoms and alkali-like ions.}\BibitemShut {Stop}%
\end{thebibliography}

\onecolumngrid
\pagebreak



\section*{Supplementary material}

# Supplemental Material for
# Super-radiant and Sub-radiant Cavity Scattering by Atom Arrays

Zhenjie Yan, Jacquelyn Ho, and Yue-Hui Lu
*Department of Physics, University of California, Berkeley, California 94720 and*
*Challenge Institute for Quantum Computation, University of California, Berkeley, California 94720*

Stuart J. Masson and Ana Asenjo-Garcia
*Department of Physics, Columbia University, New York, NY 10027*

Dan M. Stamper-Kurn\*
*Department of Physics, University of California, Berkeley, California 94720*
*Challenge Institute for Quantum Computation, University of California, Berkeley, California 94720 and*
*Materials Sciences Division, Lawrence Berkeley National Laboratory, Berkeley, California 94720*

## I. EXPERIMENTAL METHODS

The one-dimensional optical tweezer array is formed by multiple beams at a wavelength of 808 nm generated by an acousto-optic deflector (AOD), similar to the method described in Ref. [S1]. The spacing between the tweezers is controlled by the frequency differences between the radio-frequency tones driving the AOD. The center-of-mass position of the tweezer array in the probe-cavity plane ($x$-$y$ plane) is controlled using a piezoactuated mirror. The AOD is mounted on a motorized rotational stage for precise control of the angle of the tweezer array. For the two-atom cavity scattering data shown in Fig. 2(a) in the main text, we probabilistically load atoms in two static tweezers and post-select data with both sites loaded. For the rest of the data, a deterministic loading procedure is used to create atom arrays with the desired atom number and spacing. Atoms are initially loaded into an array of 16 optical tweezers, followed by polarization gradient cooling (PGC). A fluorescence image determines the site occupation after parity projection caused by laser cooling. Subsequently, the empty tweezers are diminished, and the occupied tweezers are rearranged to an array with the desired geometry [S2]. When the loaded atom number exceeds the target, the tweezer sites containing extra atoms are diminished after rearrangement. Finally, a second fluorescence image verifies the success of the tweezer rearrangement process. In our experiment, the success rate of generating an array of $N = 8$ atoms from 16 initial tweezers is about 50%. The main limitation on the achievable atom number in this work is the laser power for tweezer array. During the preparation of this manuscript, higher tweezer-trap laser power was employed to enable loading of as many as $N = 20$ atoms deterministically within the cavity.

Controlling the collective coupling between the cavity and the atom array requires precise calibration of the array orientation and interatomic spacing with respect to both the standing-wave light used to probe the array and the standing-wave cavity mode. The cavity photon number emitted by a single atom $n_1(x,y) \propto |\eta(x,y)|^2$ forms a 2D square lattice with a period of $\lambda/2 = 390$ nm in the $x$-$y$ plane and can serve as a positional reference for each atom [Fig. S1(a)]. To achieve a precise alignment between the tweezer array and the wavefront of the side probe light (along the $x$-axis), we compare the photon scattering rates from two single atoms while scanning both tweezers along the $y$-axis (propagation axis of the probe light), as shown in Fig. S1(b). By fine-tuning the angle of the acousto-optic deflector (AOD) using a motorized rotational stage, we phase-match the photon scattering rates along the side probe standing wave from the two atoms. This measurement is first carried out at an atom spacing $d = 4\lambda$ and repeated at increasing $d$ to improve angle alignment precision. We achieve a phase matching within $0.1\pi$ for photon scattering rate between two atoms $d = 40\lambda$ apart, corresponding to an angle alignment accuracy better than $6 \times 10^{-4}$ rad. Similarly, the interatomic spacing is calibrated by comparing the phase difference between the photon scattering rates from two single atoms when they are scanned along the cavity axis ($x$-axis). We note a gradual spatial drift of both the cavity and probe standing waves, amounting to less than 200 nm over a 3-hour data collection period. This timescale significantly exceeds the 5-second experimental iteration time, which includes scattering rate measurements at up to 16 distinct center-of-mass positions for the tweezer array. This standing wave drift is mitigated through feedback to the positions of tweezers upon periodic calibration of the 2D standing waves. The contrast of the photon scattering signal from a single atom moving along the cavity or probe axis also provides an independent measurement of the atomic position

---

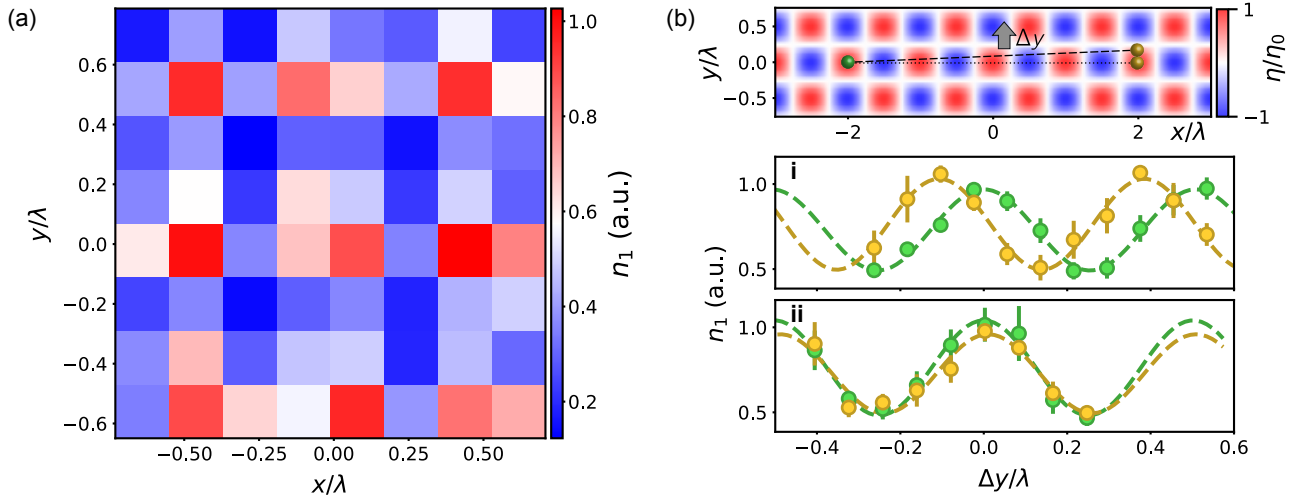

Figure S1. (a) The cavity photon number $n_1(x,y)$ scattered by a single atom positioned at position $(x,y)$. (b) Measuring the orientation of the tweezer array with respect to the probe standing wave. The single-atom emission $n_1(\Delta y)$ (green and yellow circles) from two tweezers is measured when the array is translated along the $y$-axis by $\Delta y$. The array is misaligned (i) or aligned (ii) with the probe wavefront when the measured $n_1(\Delta y)$ from the two tweezers is mismatched or matched.

fluctuations in the 2D plane [see Eq. (S16)]. A consistent contrast of $C = 30(10)\%$ is observed along both directions; the error bar here represents the overall drift during several months of data taking. The measured contrast here corresponds to rms fluctuations of the atomic position within the tweezer trap, during single exposures of the probe light, of $\sigma = 100(14)$ nm in the $x-y$ plane, where the single atom density profile is $\rho(\mathbf{r}) = \frac{1}{2\pi\sigma^2}\exp\left[-\frac{1}{2}(\mathbf{r}-\mathbf{r_0})^2/\sigma^2\right]$.

The steady state photon number $n_N = R/(2\kappa P)$ during the collective cavity scattering measurement is calculated with the photon detection rate $R$ on the single photon counting module, the overall cavity photon detection efficiency $P = 25\%$, and the cavity photon decay rate $2\kappa$. During the cavity scattering measurement near the magic detuning, PGC is applied to compensate for heating induced by photon recoils since the atoms experience a vanishing tensor ac Stark shift from the probe light. When measuring the cavity photon signal close to the atomic resonance, we cycle between a short probe light pulse of $50\,\mu s$ and a PGC of $1.7\,ms$. An additional repump beam, with an intensity well above the saturation intensity, resonantly drives the $F = 1 \rightarrow F' = 2$ transition and pumps atoms that have fallen into the $F = 1$ states back to the $F = 2$ states. The measurements are conducted without a bias magnetic field. The maximum probe Rabi frequency of the side probe standing wave on the $F = 2 \rightarrow F' = 3$ cycling transition is $\Omega_0 = 2\pi \times 13.0\,MHz$ for the measurements at the magic detuning, $\Omega_0 = 2\pi \times 4.2\,MHz$ for $\Delta_{ca} = -2\pi \times 38\,MHz$, and $\Omega_0 = 2\pi \times 2.0\,MHz$ for $\Delta_{ca} = \pm 2\pi \times 19\,MHz$.

## II. THEORY OF CAVITY PHOTON SCATTERING

For an array of $N$ spatially fixed two-level atoms, the Rayleigh scattering process from the probe light through the atoms into a single cavity mode can be described by the following Hamiltonian $\hat{H}$:

$$\hat{H}/\hbar = -\Delta_{pc}\hat{a}^\dagger\hat{a} + \sum_{i=1}^{N}\left\{-\frac{\Delta_{ca}}{2}\hat{\sigma}_{z,i} + \left[\frac{\Omega(y_i)}{2} + g(x_i)\hat{a}\right]\hat{\sigma}_i^+ + \left[\frac{\Omega(y_i)}{2} + g(x_i)\hat{a}^\dagger\right]\hat{\sigma}_i^-\right\}, \tag{S1}$$

where $\hat{a}$ ($\hat{a}^\dagger$) is the cavity photon annilation (creation) operator, $\hat{\sigma}_{z,i}$ is the Pauli $z$ matrix for the $i$th two-level emitter, and $\hat{\sigma}_i^+$ ($\hat{\sigma}_i^-$) is the raising (lowering) operator for that emitter. Here we assume $\Omega$ and $g$ are both real. The jump operator $\hat{L}_i = \sqrt{2\gamma}\hat{\sigma}_i^-$ describes the spontaneous decay of an atom emitting a photon to non-cavity modes, and the jump operator $\hat{L}_c = \sqrt{2\kappa}\hat{a}$ describes the leakage of cavity photons. The time evolution of a quantum operator $\hat{X}$ in the Heisenberg picture is governed by the master equation:

$$\partial_t\hat{X} = \frac{i}{\hbar}[\hat{H},\hat{X}] + \sum_{j\in\{1,\ldots,N,c\}}\left(\hat{L}_j^\dagger\hat{X}\hat{L}_j - \frac{1}{2}\hat{L}_j^\dagger\hat{L}_j\hat{X} - \frac{1}{2}\hat{X}\hat{L}_j^\dagger\hat{L}_j\right). \tag{S2}$$

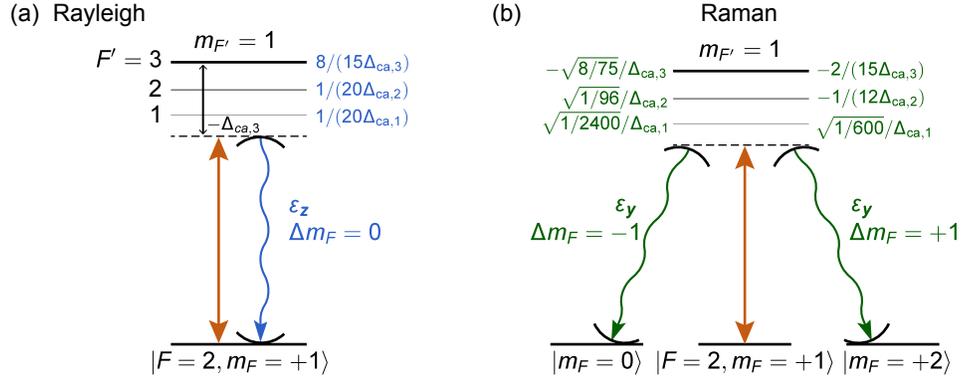

Figure S2. Rayleigh (a) and Raman (b) scattering from the side probe light (brown) into the two cavity modes with $z$- (blue) and $y$- (green) linear polarization through a single atom in the $|F = 2, m_F = +1\rangle$ state. Multiple hyperfine states in the excited manifold contribute to the two-photon scattering process. The numbers beside the intermediate states indicate the relative amplitudes of the two-photon scattering processes through that state to a final state with $\Delta m_F = 0$ (Rayleigh) or $\Delta m_F = \pm 1$ (Raman), with the assumption of $|\Delta_{\text{ca},F'}| \gg \gamma$.

The steady state of the driven dissipative system is reached when the time derivatives of the quantum operators all equal zero:

$$\partial_t \hat{\sigma}_i^- = i \left\{ \Delta_{\text{ca}} \hat{\sigma}_i^- + \left[ \frac{\Omega(y_i)}{2} + g(x_i) \hat{a} \right] \hat{\sigma}_{z,i} \right\} - \gamma \hat{\sigma}_i^- = 0. \tag{S3}$$

$$\partial_t \hat{a} = i \left[ \Delta_{\text{pc}} \hat{a} - \sum_i g(x_i) \hat{\sigma}_i^- \right] - \kappa \hat{a} = 0, \tag{S4}$$

In the dispersive coupling regime where $|\Delta_{\text{ca}}| \gg \gamma, \Omega_0, g_0$, the population in the atomic excited state is negligible and the steady state value of $\hat{\sigma}_{z,i}$ is $\bar{\sigma}_{z,i} \simeq -1$. This leads to the steady state solution of the atomic dipole operator $\bar{\sigma}_i^-$ and photon operator $\bar{a}$:

$$\bar{\sigma}_i^- = \frac{\Omega(y_i)/2 + g(x_i)\bar{a}}{\Delta_{\text{ca}} + i\gamma}, \tag{S5}$$

$$\bar{a} = \frac{\sum_i g(x_i)\Omega(y_i)/2(\Delta_{\text{ca}} + i\gamma)}{[\Delta_{\text{pc}} - \sum_i g^2(x_i)/\Delta_{\text{ca}}] + i[\kappa + \sum_i \gamma g^2(x_i)/\Delta_{\text{ca}}^2]}. \tag{S6}$$

In the main text, the above expression for $\bar{a}$ is simplified further by dropping $+i\gamma$ term in the numerator, justified by the dispersive-regime condition $|\Delta_{\text{ca}}| \gg \gamma$.

In our experiment, the cavity scattering process is further complicated due to the presence of two degenerate polarization modes within the cavity, as well as the atom being a multilevel system. Here we choose the quantization axis to be aligned with the probe light polarization ($z$-axis). The cavity photon polarization can be decomposed into two linear polarization states along the $z$- and $y$- axes, with polarization vectors denoted as $\boldsymbol{\epsilon_z}$ and $\boldsymbol{\epsilon_y}$. The side probe light with linear ($z$-axis) polarization drives the atomic $\pi$ transitions. A Rayleigh or Raman scattering process emits a cavity photon with $\boldsymbol{\epsilon_z}$ or $\boldsymbol{\epsilon_y}$ polarization into the cavity and brings the atom back to the same state or causes a change in the projection of angular momentum $\Delta m_F = \pm 1$, respectively. Figure S2 illustrates all possible cavity scattering processes and the relative two-photon scattering amplitude from each possible intermediate state for an atom in the $|F = 2, m_F = 1\rangle$ state. The total two-photon transition amplitude for the cavity Rayleigh scattering ($\eta_{\text{Rayleigh}}$) and Raman scattering ($\eta_{\text{Raman}}$) with $\Delta m_F$ by an atom initially in $|m_F\rangle$ state at spatial location $(x, y)$ is:

$$\eta_{\text{Rayleigh}}(x, y, m_F) = \sum_{F'} \frac{g(x, m_F, F', \Delta_m = 0)\Omega(y, m_F, F', \Delta_m = 0)}{2(\Delta_{\text{ca},F'} + i\gamma)}, \tag{S7}$$

$$\eta_{\text{Raman}}(x, y, m_F, \Delta m_F) = \sum_{F'} \frac{g(x, m_F + \Delta m_F, F', \Delta_m = -\Delta m_F)\Omega(y, m_F, F', \Delta_m = 0)}{2(\Delta_{\text{ca},F'} + i\gamma)}; \tag{S8}$$

where $g(x, m_F, F', \Delta_m)$ and $\Omega(x, m_F, F', \Delta_m)$ are coupling amplitudes between the ground state $|m_F\rangle$ state and the excited state $|F', m_{F'} = m_F + \Delta_m\rangle$ through the cavity mode and probe light with $z$- ($\Delta_m = 0$) or $y$- ($\Delta_m = \pm 1$) linear polarization, and $\Delta_{\text{ca},F'}$ is the cavity detuning from the $F'$ excited state.

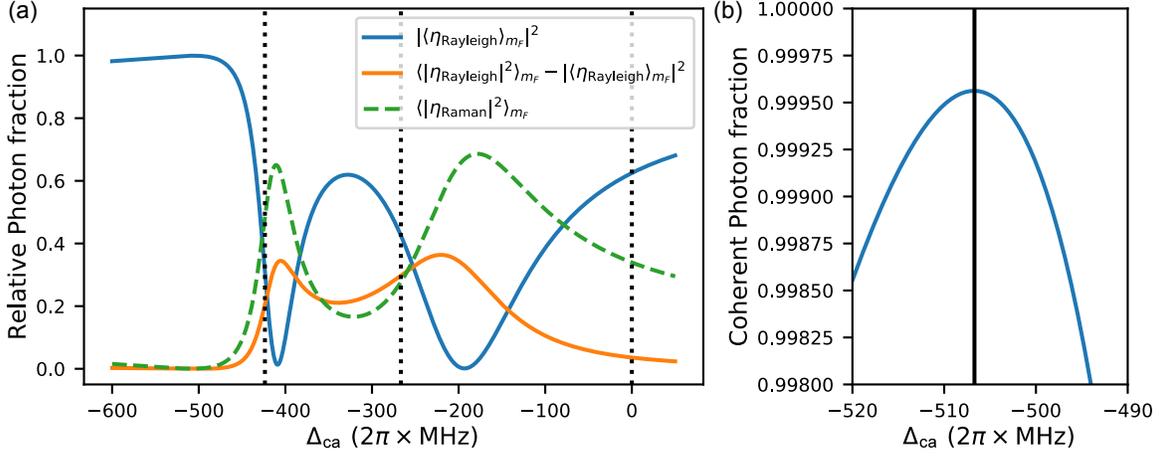

Figure S3. (a) The fraction of cavity photons generated by a single atom through Rayleigh and Raman scattering processes at a specific cavity-atom detuning $\Delta_{\text{ca}}$. The solid blue (orange) line is the fraction of coherent (incoherent) photons with $z$-polarization generated by Rayleigh scattering. The dashed green line is the incoherent Raman scattered photons with $y$-polarization. The vertical dotted lines (from left to right) indicate the resonance frequency for the $F' = 1, 2, 3$ excited states. (b) A maximum in the total coherent photon fraction occurs at the magic detuning $\Delta_{\text{ca}} = -2\pi \times 506.7$ MHz (black vertical line).

As stated in the main text, there is a distinction between Rayleigh and Raman scattering in terms of collective effects on the scattering rate. We repeat the argument here with more detail for greater clarity.

Consider the $N$-atom array is initially found with all atoms in the ground state $F = 2$ manifold, in the quantum state $|m_{F,1}, m_{F,2}, \ldots m_{F,N}\rangle$ where $m_{F,i}$ indicates the $m_F$ magnetic quantum number for atom $i$. We assume that all atoms are in a well defined magnetic quantum state, as coherence among the ground states should not develop with a weak probe due to the imprecise initial state preparation, magnetic field noises, and state-dependent scattering. Now consider a single scattering event in which a side probe photon is absorbed, and a cavity photon is emitted. Rayleigh scattering leads to a transformation of the state of the system as

$$|m_{F,1}, m_{F,2}, \ldots m_{F,N}\rangle |0\rangle_{c,z} |0\rangle_{c,y} \xrightarrow{\text{Rayleigh}} |m_{F,1}, m_{F,2}, \ldots m_{F,N}\rangle |1\rangle_{c,z} |0\rangle_{c,y} \sum_i \eta_i^{\text{Rayleigh}}. \tag{S9}$$

One sees that the final quantum state is unique, with the atomic quantum state unchanged while the $\boldsymbol{\epsilon_z}$ cavity mode is promoted from the vacuum state $|0\rangle_{c,z}$ to the one-photon state $|1\rangle_{c,z}$. The overall cavity field intensity is then proportional to $|\sum_i \eta_i^{\text{Rayleigh}}|^2$. The probability amplitudes for emission from each of the atoms interfere, either constructively or destructively, allowing for either super- or sub-radiant emission, respectively. In the case of super-radiant emission ($\eta_i^{\text{Rayleigh}} = \eta_j^{\text{Rayleigh}}$) $\forall\ i, j$, this leads to a cavity photon number that scales as $N^2$.

Raman scattering leads to a transformation of the state of the system as

$$|m_{F,1}, m_{F,2}, \ldots m_{F,N}\rangle |0\rangle_{c,z} |0\rangle_{c,y} \xrightarrow{\text{Raman}} \sum_i \eta_i^{\text{Raman}} |m_{F,1}, m_{F,2}, \ldots m_{F,i} \pm 1 \ldots m_{F,N}\rangle |0\rangle_{c,z} |1\rangle_{c,y}. \tag{S10}$$

In the case of Raman scattering, the final atomic state is dependent on which atom scattered the photon and so the emitted fields do not interfere. We use the short-hand notation $m_{F,i} \pm 1$ to indicate that the $i^{\text{th}}$ atom is in some superposition of the $|m_{F,i} + 1\rangle$ and $|m_F = -1\rangle$ produced by the emission into the cavity of an $\boldsymbol{\epsilon_y}$ polarized photon. The total Raman scattering rate is simply the summation of the individual scattering amplitude *in quadrature*, and is then proportional to $\sum_i |\eta_i^{\text{Raman}}|^2$.

Super-radiant bursts of inelastic Raman scattering *can* occur once coherent superpositions such as those in the final step of Eq. (S10) are preserved. However, here, we operate on the assumption that coherence of the atomic state over multiple photon emissions is not preserved, and we focus specifically on single-photon emission processes from incoherent (e.g. product state) ground states of the atom array.

In Fig. S2, we present the relative intensity of Rayleigh and Raman scattered cavity light produced by a single atom. Here, we assume that incoherent processes cause the atom to be distributed initially equally among all $m_F$ states in the $F = 2$ ground state. The total photon number from Rayleigh (Raman) processes is proportional to $\langle |\eta_{\text{Rayleigh}}|^2 \rangle_{m_F}$ ($\langle |\eta_{\text{Raman}}|^2 \rangle_{m_F}$), where $\langle \rangle_{m_F}$ denotes an average over all $m_F$ states and summing over $\Delta m_F = \pm 1$ for the Raman transition amplitudes. The averaged coherent amplitude of the cavity field with $z$-polarization is

proportional to $\langle \eta_{\text{Rayleigh}} \rangle_{m_F}$. The fractions of coherent and incoherent photons scattered by an unpolarized atom are shown in Fig. S3. At a magic detuning of $\Delta_{ca} = -2\pi \times 506.7$ MHz, the incoherent two-photon cavity scattering almost disappears due to destructive interference between the intermediate states, and all $m_F$ states interact almost exclusively with the $z-$polarized cavity mode with equal strengths. Notably, we find that such a magic detuning exists universally for all the alkali atoms and alkali-like ions whose hyperfine splitting is dominated by the magnetic dipole interaction [S3]. At the magic detuning, the collective cavity scattering process can be approximated as Rayleigh scattering into a single cavity mode with a two-photon scattering amplitude $\eta(x_i, y_i) = \eta_0 \cos(kx_i) \cos(ky_i)$ from each scatterer. The cavity shift and broadening with $N \simeq 10$ atoms is also negligible compared to the empty cavity width $\kappa$. The steady state cavity field at the magic detuning is thus simply:

$$\bar{a} = \frac{\sum_i \eta(x_i, y_i)}{\Delta_{\text{pc}} + i\kappa}.$$ (S11)

Finally, we further include the effect of atomic motion on the cavity scattering. At the magic detuning, the measured cavity photon number $n$ is simply a spatial average over the time-averaged density profile of each atom $\rho_i(x_i, y_i) = \frac{1}{2\pi\sigma^2} \exp\left[-\frac{(x_i - x_{i0})^2 + (y_i - y_{i0})^2}{2\sigma^2}\right]$:

$$n = \int \frac{1}{\Delta_{\text{pc}}^2 + \kappa^2} \left| \sum_i \eta(x_i, y_i) \right|^2 \prod_i [\rho_i(x_i, y_i) \mathrm{d}x_i \mathrm{d}y_i]$$

$$= \frac{1}{\Delta_{\text{pc}}^2 + \kappa^2} \left( \sum_i \langle |\eta|^2 \rangle_i + \sum_{i \neq j} \langle \eta \rangle_i^* \langle \eta \rangle_j \right);$$ (S12)

where $\langle |\eta|^2 \rangle_i = \int |\eta_0|^2 \cos^2(kx_i) \cos^2(ky_i) \rho_i(x_i, y_i) \mathrm{d}x_i \mathrm{d}y_i,$ (S13)

$$\langle \eta \rangle_i = \int \eta_0 \cos(kx_i) \cos(ky_i) \rho_i(x_i, y_i) \mathrm{d}x_i \mathrm{d}y_i.$$ (S14)

The cavity photon signal generated from a single atom centered at $(x_0, y_0)$ read as

$$n_1(x_0, y_0) = \frac{|\eta_0^2|}{\Delta_{\text{pc}}^2 + \kappa^2} \frac{(C \cos 2kx_0 + 1)(C \cos 2ky_0 + 1)}{4},$$ (S15)

where the contrast $C$ of the cavity signal with the atom scanning along the cavity or probe axis is

$$C = e^{-2k^2\sigma^2}.$$ (S16)

For an $N$ atom array with integer wavelength spacing, the spatial average over each atom is identical: $\langle \eta \rangle_i = \langle \eta \rangle$, $\langle |\eta|^2 \rangle_i = \langle |\eta|^2 \rangle$, where

$$\langle \eta \rangle = \int \eta_0 \cos(kx) \cos(ky) \frac{1}{2\pi\sigma^2} \exp\left[-\frac{(x^2 + y^2)}{2\sigma^2}\right] \mathrm{d}x \mathrm{d}y = \eta_0 e^{-k^2\sigma^2},$$ (S17)

$$\langle |\eta|^2 \rangle = \int |\eta_0|^2 \cos^2(kx) \cos^2(ky) \frac{1}{2\pi\sigma^2} \exp\left[-\frac{(x^2 + y^2)}{2\sigma^2}\right] \mathrm{d}x \mathrm{d}y = |\eta_0|^2 \left(\frac{e^{-2k^2\sigma^2} + 1}{2}\right)^2.$$ (S18)

This yields a total photon count

$$n_N = \frac{1}{\Delta_{\text{pc}}^2 + \kappa^2} \left[ N(\langle |\eta|^2 \rangle - |\langle \eta \rangle|^2) + N^2 |\langle \eta \rangle|^2 \right],$$ (S19)

which is equivalent to Eq. (2) in the main text at $\Delta_{\text{pc}} = 0$. For an array with half-integer wavelength spacing, the spatial average over each atom is $\langle \eta \rangle_i = (-1)^i \langle \eta \rangle$ and $\langle |\eta|^2 \rangle_i = \langle |\eta|^2 \rangle$, yielding a total photon count

$$n_N = \frac{1}{\Delta_{\text{pc}}^2 + \kappa^2} \left[ N(\langle |\eta|^2 \rangle - |\langle \eta \rangle|^2) + \frac{1 - (-1)^N}{2} |\langle \eta \rangle|^2 \right].$$ (S20)

The expressions in Eq. (S19) and (S20) also apply when the side probe light is replaced with a traveling wave and $\eta_{\text{tw}}(x,y) = \eta_0 \cos(kx)\exp(iky)$. In such configuration, the coherent scattering rate from an atom on the cavity antinode is identical to that with standing-wave probe light:

$$|\langle\eta_{\text{tw}}\rangle|^2 = |\langle\eta\rangle|^2 = |\eta_0|^2 e^{-2k^2\sigma^2}, \tag{S21}$$

while the total emission rate $\langle|\eta_{\text{tw}}|^2\rangle$, hence the incoherent emission rate $\langle|\eta_{\text{tw}}|^2\rangle - |\langle\eta_{\text{tw}}\rangle|^2$, is higher:

$$\langle|\eta_{\text{tw}}|^2\rangle = |\eta_0|^2 \left(\frac{e^{-2k^2\sigma^2}+1}{2}\right) \geq |\langle\eta|^2\rangle. \tag{S22}$$

When the cavity frequency is close to that of one of the atomic excited states, the cavity frequency shift and broadening become nonnegligible and depend on the position of each atom. We use a Monte Carlo method to calculate the cavity photon scattering from an atom array. The spatial position $(x_i, y_i)$ of each atom is sampled with probability distribution $\rho_i(x_i, y_i)$, and the projection of atomic angular momentum $m_{Fi}$ is uniformly sampled between $m_{Fi} \in [-2, 2]$. For each instance of $\{x_i, y_i, m_{Fi}\}$, the spectral shifts $\Delta_c$ and broadenings $\Delta\kappa$ for the two linear polarization modes $\boldsymbol{\epsilon} = \boldsymbol{\epsilon_z}$ or $\boldsymbol{\epsilon_y}$ are

$$\Delta_c(\boldsymbol{\epsilon}) = \begin{cases} \sum_{i,F'} \frac{g^2(x_i, m_{Fi}, F', \Delta_m = 0)}{\Delta_{\text{ca},F'}} & \text{with } \boldsymbol{\epsilon} = \boldsymbol{\epsilon_z}, \\ \sum_{i,F',\Delta_m=\pm1} \frac{g^2(x_i, m_{Fi}, F', \Delta_m)}{\Delta_{\text{ca},F'}} & \text{with } \boldsymbol{\epsilon} = \boldsymbol{\epsilon_y}; \end{cases} \tag{S23}$$

$$\Delta\kappa(\boldsymbol{\epsilon}) = \begin{cases} \sum_{i,F'} \frac{g^2(x_i, m_{Fi}, F', \Delta_m = 0)}{\Delta_{\text{ca},F'}^2}\gamma & \text{with } \boldsymbol{\epsilon} = \boldsymbol{\epsilon_z}, \\ \sum_{i,F',\Delta_m=\pm1} \frac{g^2(x_i, m_{Fi}, F', \Delta_m)}{\Delta_{\text{ca},F'}^2}\gamma & \text{with } \boldsymbol{\epsilon} = \boldsymbol{\epsilon_y}. \end{cases} \tag{S24}$$

The cavity photon number $n(\Delta_{\text{pc}}, \boldsymbol{\epsilon})$ in $\boldsymbol{\epsilon_z}$ and $\boldsymbol{\epsilon_y}$ polarization is generated through coherent Rayleigh scattering and incoherent Raman scattering, respectively:

$$n(\Delta_{\text{pc}}, \boldsymbol{\epsilon_z}) = \left|\frac{\sum_i \eta_{\text{Rayleigh}}(x_i, y_i, m_{Fi})}{\Delta_{\text{pc}} - \Delta_c(\boldsymbol{\epsilon_z}) + i[\kappa + \Delta\kappa(\boldsymbol{\epsilon_z})]}\right|^2, \tag{S25}$$

$$n(\Delta_{\text{pc}}, \boldsymbol{\epsilon_y}) = \sum_{i,\Delta m_F=\pm1} \left|\frac{\eta_{\text{Raman}}(x_i, y_i, m_{Fi}, \Delta m_F)}{\Delta_{\text{pc}} - \Delta_c(\boldsymbol{\epsilon_y}) + i[\kappa + \Delta\kappa(\boldsymbol{\epsilon_y})]}\right|^2. \tag{S26}$$

The total emission spectrum $n_N(\Delta_{\text{pc}})$ is the sum of the spectra in both polarizations:

$$n_N(\Delta_{\text{pc}}) = n(\Delta_{\text{pc}}, \boldsymbol{\epsilon_z}) + n(\Delta_{\text{pc}}, \boldsymbol{\epsilon_y}). \tag{S27}$$

A Lorentzian fit on the averaged $n_N(\Delta_{\text{pc}})$ over all instances of $\{x_i, y_i, m_{Fi}\}$ gives the ensemble averaged spectral heights, shifts, and widths. The cavity photon number measured in a linear polarization mode $\boldsymbol{\epsilon_\theta} = \cos\theta\boldsymbol{\epsilon_z} + \sin\theta\boldsymbol{\epsilon_y}$ is

$$n_N(\Delta_{\text{pc}}, \theta) = \left|\frac{\sum_i \eta_{\text{Rayleigh}}(x_i, y_i, m_{Fi})}{\Delta_{\text{pc}} - \Delta_c(\boldsymbol{\epsilon_z}) + i[\kappa + \Delta\kappa(\boldsymbol{\epsilon_z})]}\right|^2 \cos^2\theta + \sum_{i,\Delta m_F=\pm1} \left|\frac{\eta_{\text{Raman}}(x_i, y_i, m_{Fi}, \Delta m_F)}{\Delta_{\text{pc}} - \Delta_c(\boldsymbol{\epsilon_y}) + i[\kappa + \Delta\kappa(\boldsymbol{\epsilon_y})]}\right|^2 \sin^2\theta \tag{S28}$$

averaged over all instances of $\{x_i, y_i, m_{Fi}\}$.

---



\end{document}